\documentclass[review,3p]{elsarticle}

\usepackage{lineno,hyperref}
\usepackage{booktabs}
\usepackage{multirow}
\usepackage{graphicx}
\usepackage{amsfonts,bm}
\usepackage{amssymb}
\usepackage{enumerate}
\usepackage{tabularx}
\usepackage{setspace}
\usepackage{mathrsfs}
\usepackage{diagbox}
\usepackage[table,xcdraw]{xcolor}
\modulolinenumbers[5]

\journal{Journal of \LaTeX\ Templates}

\usepackage{numcompress}\biboptions{authoryear}

\begin{document}

\begin{frontmatter}

\title{Identifying the key components in ResNet-50 for diabetic retinopathy grading from fundus images: a systematic investigation}


\author[address-a,address-b]{Yijin Huang}

\author[address-a,address-c]{Li Lin}
\author[address-a]{Pujin Cheng}
\author[address-a,address-d]{Junyan Lyu}
\author[address-b]{Roger Tam\corref{corr}}
\ead{roger.tam@ubc.ca}
\author[address-a]{Xiaoying Tang\corref{corr}}
\ead{tangxy@sustech.edu.cn}
\cortext[corr]{Corresponding authors}

\address[address-a]{Department of Electronic and Electrical Engineering, Southern University of Science and Technology, China}
\address[address-b]{School of Biomedical Engineering, The University of British Columbia, Canada}
\address[address-c]{Department of Electrical and Electronic Engineering, The University of Hong Kong, China}
\address[address-d]{Queensland Brain Institute, The University of Queensland, Australia}

\begin{abstract}
Although deep learning based diabetic retinopathy (DR) classification methods typically benefit from well-designed architectures of convolutional neural networks, the training setting also has a non-negligible impact on the prediction performance. The training setting includes various interdependent components, such as objective function, data sampling strategy and data augmentation approach. To identify the key components in a standard deep learning framework (ResNet-50) for DR grading, we systematically analyze the impact of several major components. Extensive experiments are conducted on a publicly-available dataset EyePACS. We demonstrate that (1) the DR grading framework is sensitive to input resolution, objective function, and composition of data augmentation, (2) using mean square error as the loss function can effectively improve the performance with respect to a task-specific evaluation metric, namely the quadratically-weighted Kappa, (3) utilizing eye pairs boosts the performance of DR grading and (4) using data resampling to address the problem of imbalanced data distribution in EyePACS hurts the performance. Based on these observations and an optimal combination of the investigated components, our framework, without any specialized network design, achieves the state-of-the-art result (0.8631 for Kappa) on the EyePACS test set (a total of 42670 fundus images) with only image-level labels. We also examine the proposed training practices on other fundus datasets and other network architectures to evaluate their generalizability. Our codes and pre-trained model are available at \href{https://github.com/YijinHuang/pytorch-classification}{https://github.com/YijinHuang/pytorch-classification}.
\end{abstract}


\end{frontmatter}


\section{Introduction}

Diabetic retinopathy (DR) is one of the microvascular complications of diabetes, causing vision impairments and blindness \citep{review, diabetic}. The major pathological signs of DR include hemorrhages, exudates, microaneurysms, and retinal neovascularization. The digital color fundus image is the most widely used imaging modality for ophthalmologists to screen and identify the severity of DR, which can reveal the presence of different lesions. An early diagnosis and timely intervention of DR is of vital importance in preventing patients from vision malfunction. However, due to the rapid increase in the number of patients at risk of developing DR, ophthalmologists in regions with limited medical resources bear a heavy labor-intensive burden in DR screening. As such, developing automated and efficient DR diagnosis and prognosis approaches is urgently needed to reduce the number of untreated patients and the burden of ophthalmic experts.

\begin{figure}[t]
    \centering
    \includegraphics[width=0.6\linewidth]{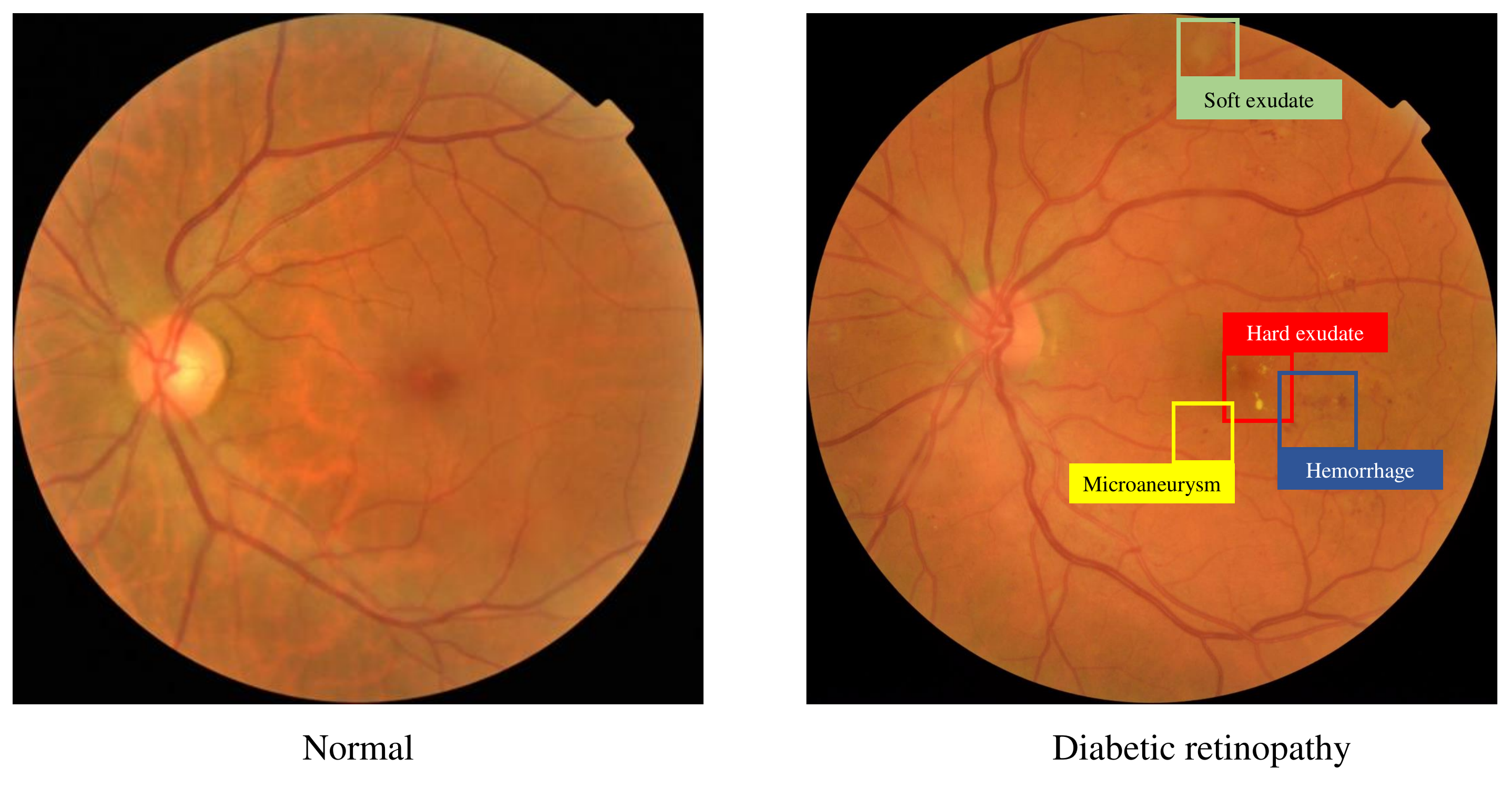}
    \label{fig:1}
    \caption{A normal fundus image (left) and a representative DR fundus image with multiple types of lesions (right).}
\end{figure}

Based on the type and quantity of lesions in fundus images, DR can be classified into five grades: 0 (normal), 1 (mild DR), 2 (moderate DR), 3 (severe DR), and 4 (proliferative DR) \citep{sustech}. Red dot-shaped microaneurysms are the first visible sign of DR, and their presence indicates a mild grade of DR. Red lesions (e.g., hemorrhages) and yellow-white lesions (e.g., hard exudates and soft exudates) have various types of shapes, from tiny points to large patches. A larger amount of such lesions indicate severer DR grading. Neovascularization, the formation of new retinal vessels in the optic disc or its periphery, is a significant sign of proliferative DR. Fig.~\ref{fig:1} shows examples of fundus images with different types of lesions.

In recent years, deep learning based methods have achieved great success in the field of computer vision. With the capability of highly representative feature extraction, convolutional neural networks (CNNs) have been proposed to tackle different tasks. They have also been widely used in the medical image analysis realm \citep{seg, miadr, tripleA, boundary, lin2021bsda}. In DR grading, \cite{cvdr} adopts a pre-trained CNN as a feature extractor and re-trains the last fully connected layer for DR detection. Given that lesions are important guidance in DR grading \citep{huang2021lesion}, Attention Fusion Network \citep{afn} employs a lesion detector to predict the probabilities of lesions and proposes an information fusion method based on an attention mechanism to identify DR. Zoom-in-net \citep{zoom} consists of three sub-networks which respectively localize suspicious regions, analyze lesion patches and classify the image of interest. To enhance the capability of a standard CNN, CABNet \citep{cabnet} introduces two extra modules, one for exploring region-wise features for each DR grade and one for generating attention feature maps.

It can be observed that recent progress in automatic DR grading is largely attributed to carefully designed model architecture. Nevertheless, the task-specific designs and specialized configurations may limit their transferability and extensibility. Other than model architecture, the training setting is also a key factor affecting the performance of a deep learning method. A variety of interdependent components are typically involved in a training setting, including the design of configurations (e.g., preprocessing, loss function, sampling strategy, and data augmentation) and empirical decisions of hyper-parameters (e.g., input resolution, learning rate, and training epochs). Proper training settings can benefit automatic DR grading, while improper ones may damage the grading performance. However, the importance of the training setting has been overlooked or received less attention in the past few years, especially in the DR grading field. In computer vision, there have been growing efforts in improving the performance of deep learning methods by refining the training setting rather than the network architecture. For example, \cite{bag} boosts ResNet-50's \citep{resnet} top-1 validation accuracy from 75.3\% to 79.29\% on ImageNet \citep{imagenet} by applying numerous training procedure refinements. \cite{yolov4} examines combinations of training configurations such as batch-normalization and residual-connection, and utilizes them to improve the performance of object detection. In the biomedical domain, efforts in this direction have also emerged. For example, \cite{nnu} proposes an efficient deep learning-based segmentation framework for biomedical images, namely nnU-Net, which can automatically and optimally configure its own setting including preprocessing, training and post-processing. In such context, we believe that refining the training setting has a great potential in enhancing the DR grading performance.

\begin{figure}[t]
    \centering
    \includegraphics[width=\linewidth]{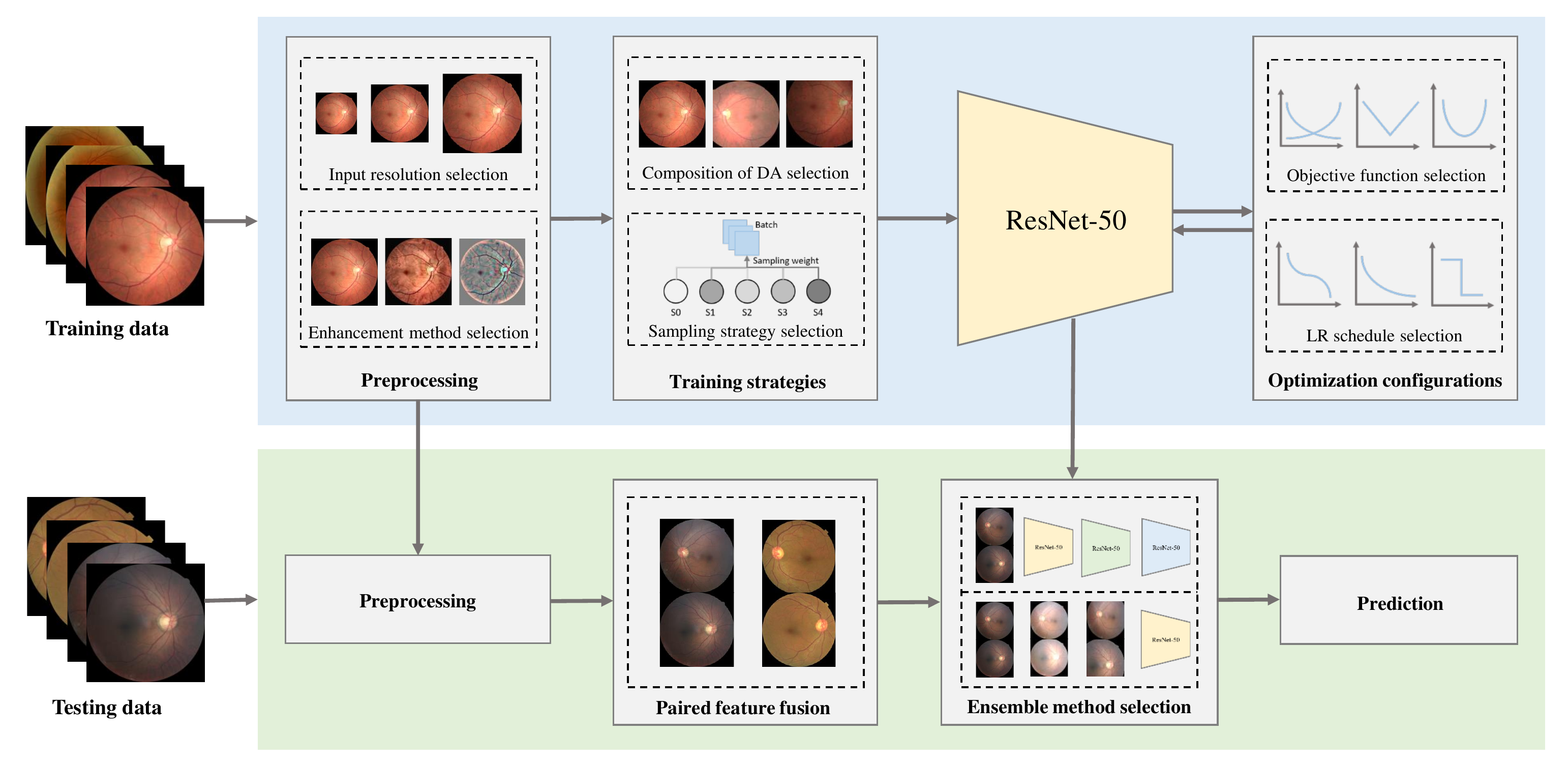}
    \caption{Components analyzed in our deep learning-based DR grading framework. The evaluation process of a framework can be divided into two parts: training (top) and testing (bottom). In the training phase, we first fix the architecture of the selected network (ResNet-50). Then we examine a collection of designs with respect to the training setting including preprocessing (image resizing and enhancement), training strategies (compositions of data augmentation (DA) and sampling strategies) and optimization configurations (objective functions and learning rate (LR) schedules). In the testing phase, we apply the same preprocessing as in the training phase and employ paired feature fusion to make use of the correlation between the two eyes (the training step of the fusion network is omitted in this figure). Then, we select the best ensemble method for the final prediction.}
    \label{fig:pipeline}
\end{figure}

In this work, we systematically analyze the influence of several major components of a standard DR classification framework and identify the key elements in the training setting for improving the DR grading performance. We then evaluate these training practices on multiple datasets and network architectures, with a goal of analyzing their generalizability across both datasets and network architectures. The components analyzed in our work are shown in Fig.~\ref{fig:pipeline}. The main contributions of this work can be summarized as follows:

\begin{itemize}
    \item We examine a collection of designs with respect to the training setting and evaluate them on the most challenging and largest publicly-available fundus image dataset, EyePACS \footnote{https://www.kaggle.com/c/diabetic-retinopathy-detection}. We analyze and illustrate the impact of each component on the DR grading performance to identify the core ones.

    \item We adopt ResNet-50 \citep{resnet} as the backbone and achieve a quadratically-weighted Kappa of 0.8631 on the EyePACS test set, which outperforms many specifically-designed state-of-the-art methods, with only image-level labels. With the plain ResNet-50, our framework can serve as a strong, standardized, and scalable DR grading baseline. In other words, other types and directions of most methodological improvements and modifications can be easily incorporated into our framework to further improve the DR grading performance.

    \item We evaluate the proposed training practices on two external retinal fundus datasets and six popular network architectures. Consistent and similar observations on multiple datasets and across different network architectures validate the generalizability and robustness of the proposed training setting refinements and the importance of the identified components in deep learning-based methods for DR grading.

    \item We emphasize that the superior performance of our framework is not achieved by a new network architecture, a new objective function nor a new scheme. The key contribution of this work, in a more generalizable sense, is that we outline another direction to improve the performance of deep learning methods for DR grading and highlight the importance of training setting refinements in developing deep learning based pipelines. This may also shed new insights into other related fields.
\end{itemize}

The remainder of this paper is organized as follows. Section~\ref{sec:2} describes the details of our baseline framework, the default training setting, and the evaluation protocol. Descriptions of the investigated components in the training setting are presented in section~\ref{sec:3}. Extensive experiments are conducted in section~\ref{sec:4} to evaluate the DR grading performance, the influence of each refinement and the generalizability of the proposed practices. Discussion and conclusion are respectively provided in section~\ref{sec:5} and section~\ref{sec:6}.

\begin{figure}[t]
    \centering
    \includegraphics[width=0.7\linewidth]{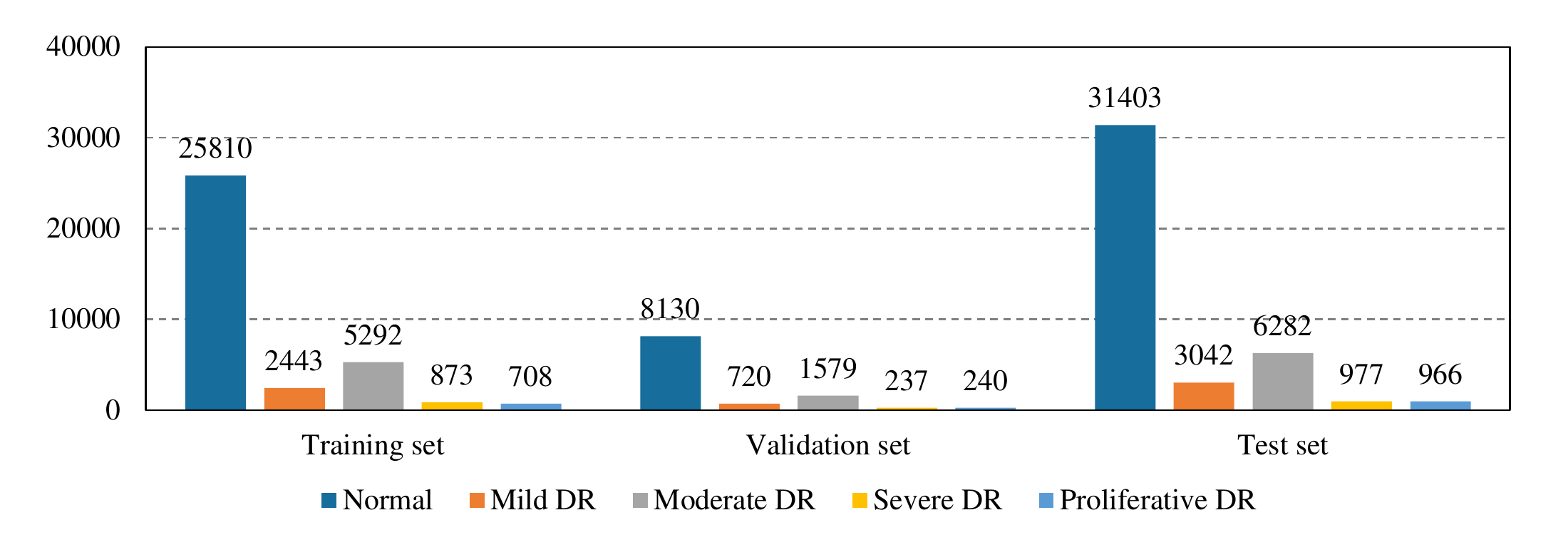}
    \caption{The imbalanced class distribution of EyePACS.}
    \label{fig:cd}
\end{figure}

\section{Method}
\label{sec:2}
\subsection{Dataset}
\label{dataset}
\textbf{EyePACS:} The EyePACS dataset is the largest publicly-available DR grading dataset released in the Kaggle DR grading competition, consisting of 88702 color fundus images from the left and right eyes of 44351 patients. Images were officially split into 35126/10906/42670 fundus images for training/validation/testing. According to the severity of DR, they have also been divided by ophthalmologists into the aforementioned five grades. The fundus images were acquired under a variety of conditions and from different imaging devices, resulting in variations in image resolution, aspect ratio, intensity, and quality \citep{cheng2021secret}. As shown in Fig.~\ref{fig:cd}, the class distribution of EyePACS is extremely imbalanced, wherein DR fundus images are dramatically less than normal images. In this work, the evaluation of each component is mainly performed on EyePACS.

\textbf{Messidor-2:} A total of 1748 fundus images with five-grade annotations and eye pairing are provided in the Messidor-2 dataset \citep{messidor}. We randomly split the dataset into 1042/176/522 fundus images for training/validation/testing. The main challenge of this dataset lies in the limited number of images for training, and thus we employ this dataset to evaluate the generalization ability of the proposed training practices.

\textbf{DDR:} The DDR dataset \citep{ddr} consists of 13673 fundus images with six-class annotations (five DR grades and another "ungradable" class). All ungradable images are excluded, ending up with 6320/2503/3759 for training/validation/testing.

\label{sec:2.1}
\subsection{Baseline setting}
We first specify our baseline for DR grading. In the preprocessing step, for each image, we first identify the smallest rectangle that contains the entire field of view and use the identified rectangle for cropping. After that, we resize each cropped image into $224 \times 224$ squares and rescale each pixel intensity value into [0, 1]. Next, we normalize the RGB channels using z-score transformations with the mean and the standard deviations obtained from the entire preprocessed training set. Common random data augmentation operations including horizontal flipping, vertical flipping, and rotation described in section~\ref{sec:da} are performed during training.

ResNet-50 is a widely used architecture in the field of deep learning. It has been adopted as a referent architecture for most analyses of training practices \citep{wightman2021resnet, yun2019cutmix, cubuk2020randaugment}. Therefore, in this work, ResNet-50 is empolyed as our baseline model for analyzing different components. We adopt the SGD optimizer with an initial learning rate of 0.001 and Nesterov Accelerated Gradient Descent \citep{nesterov} with a momentum factor of 0.9 to train the network. A weighted decay of 0.0005 is applied for regularization. Convolutional layers are initialized with parameters obtained from a ResNet-50 pre-trained on the ImageNet dataset \citep{imagenet} and the fully connected layer is initialized using He's initialization method \citep{heini}. We train the model for 25 epochs with a mini-batch size of 16 on a single NVIDIA RTX TITAN. All codes are implemented in PyTorch \citep{pytorch}. If not specified, all models are trained with a fixed random seed for fair comparisons. The model having the highest metric on the validation set is selected for testing.

\subsection{Evaluation metric}
The DR grading performance is evaluated using the quadratically-weighted Kappa $\kappa$ \citep{Kappa}, which is an officially-used metric in the Kaggle DR grading competition. In an ordinal multi-class classification task, given an observed confusion matrix $o$ and an expected matrix $e$, $\kappa$ measures their agreement by quadratically penalizing the distance between the prediction and the ground truth,
\begin{equation}
    \kappa = 1 - \frac{\sum_i^C\sum_j^C w_{ij}o_{ij}}{\sum_i^C\sum_j^C w_{ij}e_{ij}},
\end{equation}
where $C$ denotes the total number of classes, $w$ is a quadratic weight matrix, and subscripts $i$ and $j$ respectively denote the row and column indices of the matrices. The weight $w_{ij}$ is defined as $\frac{(i - j)^2}{(C - 1)^2}$. $\kappa$ ranges from $-1$ to $1$, with -1 and 1 respectively indicate total disagreement and complete agreement.
\section{Training setting components}
\label{sec:3}

\subsection{Input resolution}
The resolution of the input image has a direct impact on the DR grading performance. Generally, ResNet-50 is designed for images of $224 \times 224$ input resolution \citep{resnet}. In ResNet-50, a convolution layer with a kernel size of $7 \times 7$ and a stride of $2$ followed by a max-pooling layer is applied to dramatically downsample the input image first. Therefore, using images with very small input resolution may lose key features for DR grading, such as tiny lesions. In contrast, a network fed with large resolution images can extract more fine-grained and dense features at the cost of a smaller receptive field and a higher computational cost. In this work, a range of resolutions is evaluated to identify the trade-off.

\subsection{Loss function}
\label{sec:loss}
The objective function plays a critical role in deep learning. Let $D = \{(x_i, y_i), i = 1,...,N\}$ denote the training set, where $x_i$ is the input image and $y_i$ is the corresponding ground truth label. There are a variety of objective functions that can be used to measure the discrepancy between the predicted probability distribution $\hat{y}_i$ and the ground truth distribution $\tilde{y}_i$ (one-hot encoded $y_i$) of the given label.

\subsubsection{Cross-entropy loss}
The cross-entropy loss is the most commonly used loss function for classification tasks, which is the negative log-likelihood of a Bernoulli or categorical distribution,

\begin{equation}
    CE(\tilde{y}, \hat{y}) = -{\frac{1}{N}} \sum_{i=1}^N \tilde{y}_i \log(\hat{y}_i).
\end{equation}

\subsubsection{Focal loss}
The focal loss was initially proposed in RetinaNet \citep{focal}, which introduces a modulating factor into cross-entropy to down-weigh the loss of well-classified samples, giving more attention to challenging and misclassified ones. The focal loss is widely used to address the class imbalance problem in training deep neural networks. As mentioned before, EyePACS is an extremely imbalanced dataset with the number of images per class ranges from 25810 to 708. Therefore, the focal loss is applied for better feature learning with samples from the minority classes. The focal loss is defined as

\begin{equation}
    FL(\tilde{y}, \hat{y}) = -{\frac{1}{N}} \sum_{i=1}^N \tilde{y}_i (1 - \hat{y}_i)^\gamma \log(\hat{y}_i),
\end{equation}
where $\gamma$ is a hyper-parameter. When the predicted probability $\hat{y}_i$ is small, the modulating factor $(1 - \hat{y}_i)^\gamma$ is close to 1. When $\hat{y}_i$ is large, this factor goes to 0 to down-weigh the corresponding loss.

\subsubsection{Kappa loss}

The quadratically-weighted Kappa is sensitive to disagreements in marginal distributions, whereas cross-entropy loss does not take into account the distribution of the predictions and the magnitude of the incorrect predictions. Therefore, the soft Kappa loss \citep{de2018weighted, Kappaloss} based on the Kappa metric is another common choice for training the DR grading model,

\begin{equation}
    KL(y, \hat{y}) = 1 - \frac{o(y, \hat{y})}{e(y, \hat{y})},
\end{equation}
\begin{equation}
    o(y, \hat{y}) = \sum_{i, n} \frac{(y_i - n)^2}{(C - 1)^2} \hat{y_i},
\end{equation}
\begin{equation}
    e(y, \hat{y}) = \sum_{m,n} \frac{(m - n)^2}{(C - 1)^2} (\sum_i \mathbb{I}_{[n = y_i]})(\sum_j \hat{y}_{j,m}),
\end{equation}
where $C$ is the number of classes, $\hat{y}_{j,k}$ $(k \in [1, C])$ is the predicted probability of the $k$-th class of $\hat{y}_i$ and $\mathbb{I}_{[n = y_i]}$ is an indicator function equaling to 1 if $n = y_i$ and otherwise 0. As suggested by a previous work \citep{Kappaloss}, combining the Kappa loss with the standard cross-entropy loss can stabilize the gradient at the beginning of training to achieve better prediction performance.

\subsubsection{Regression loss}
In addition to Kappa loss, the regression loss also provides a penalty to the distance between prediction and ground truth.  When a regression loss is applied, the softmax activation of the fully connected layer is removed and the output dimension is set to be 1 to produce a prediction score $\bar{y}_i$ for the DR grade. Three regression loss functions are considered in this work, namely L1 loss (Mean Absolute Error, MAE), L2 loss (Mean Square Error, MSE), and smooth L1 loss (SmoothL1), which are respectively defined as

\begin{equation}
    \textrm{MAE}(y_i, \bar{y}_i) = {\frac{1}{N}} \sum_{i=1}^N |y_i - \bar{y}_i|,
\end{equation}
\begin{equation}
    \textrm{MSE}(y_i, \bar{y}_i) = {\frac{1}{N}} \sum_{i=1}^N (y_i - \bar{y}_i)^2,
\end{equation}
\begin{equation}
    \textrm{SmoothL1}(y_i, \bar{y}_i)= \left\{
    \begin{array}{ll}
        0.5(y_i-\bar{y}_i)^2,   & \textrm{if } |y_i-\bar{y}_i|<1 \\
        |y_i-\bar{y}_i|-0.5,    & \textrm{otherwise.}
    \end{array}
    \right .
\end{equation}

In the testing phase, the prediction scores are clipped to be between [0, 4] and then simply rounded to integers to serve as the finally predicted grades.

\subsection{Learning rate schedule}
The learning rate is important in gradient descent methods, which has non-trivial impact on the convergence of the objective function. However, the optimal learning rate may vary at different training phases. Therefore, a learning rate schedule is widely used to adjust the learning rate during training. Multiple-step decaying, exponential decaying, and cosine decaying \citep{sgdr} are popular learning rate adjustment strategies in deep learning. Specifically, the multiple-step decaying schedule decreases the learning rate by a constant factor at specific training epochs. The exponential decaying schedule exponentially decreases the learning rate by $\gamma$ at every epoch, namely

\begin{equation}
    \eta_t = \gamma^t \eta_0,
\end{equation}
where $\eta_t$ is the learning rate at epoch $t$. A typical choice of $\gamma$ is $0.9$. The cosine decaying schedule decreases the learning rate following the cosine function. Given a total number of training epochs $T$, the learning rate in the cosine decaying schedule is defined as

\begin{equation}
    \eta_t = \frac{1}{2}\left( 1 + \cos\left( \frac{t\pi}{T} \right) \right) \eta_0.
\end{equation}

The setting of the cosine decaying schedules is independent of the number of epochs, making them more flexible than other schedules.

\subsection{Composition of data augmentation}
\label{sec:da}
\begin{figure}[t]
    \centering
    \includegraphics[width=\linewidth]{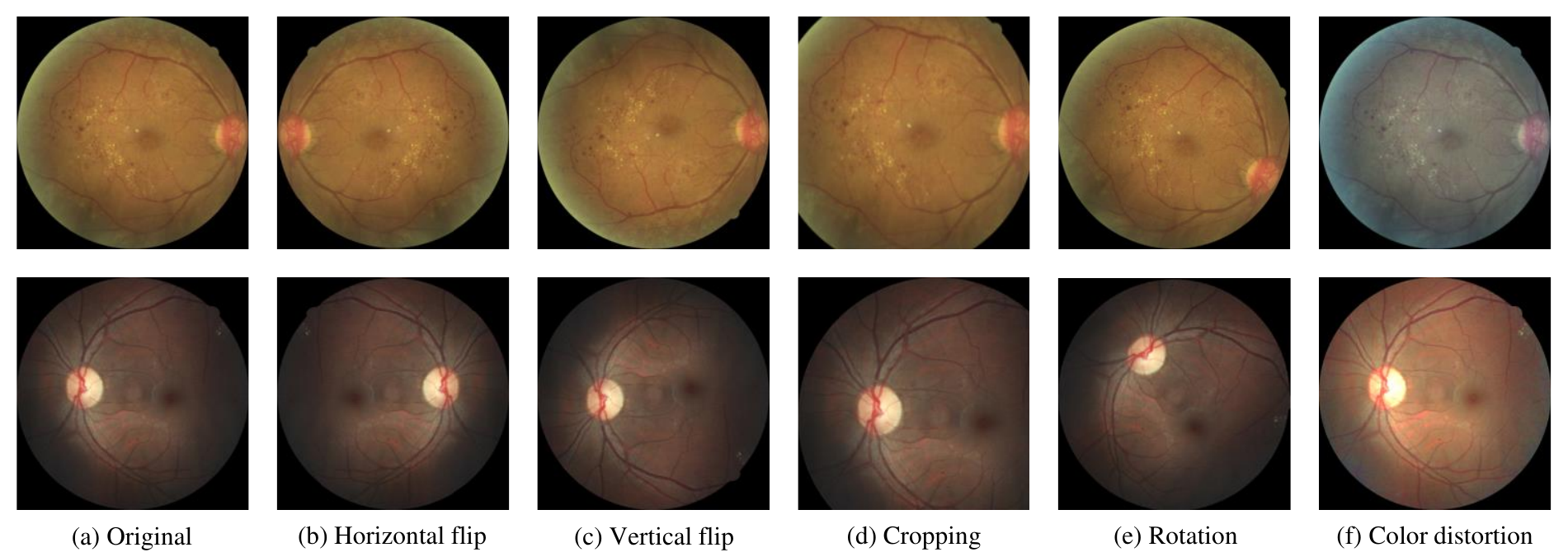}
    \caption{Illustration of common data augmentation operations.}
    \label{fig:da}
\end{figure}
Applying online data augmentation during training can increase the distribution variability of the input images to improve the generalization capacity and robustness of a model of interest. To systematically study the impact of the composition of data augmentation on DR grading, as shown in Fig.~\ref{fig:da}, various popular augmentation operations are considered in this work. For geometric transformations, we apply horizontal and vertical flipping, random rotation, and random cropping. For color transformations, color distortion is a common choice, including adjustments of brightness, contrast, saturation, and hue. Moreover, Krizhevsky color augmentation \citep{alexnet} is evaluated in our experiments, which has been suggested to be effective by the group that ranked the third place in the Kaggle DR grading competition \citep{oO}.

For the cropping operation, we randomly crop a rectangular region the size of which is randomly sampled in [1/1.15, 1.15] times the original one and the aspect ratio is randomly sampled in [0.7, 1.3], and then we resize this region back to be of the original size. Horizontal and vertical flipping is applied with a probability of 0.5. The color distortion operation adjusts the brightness, contrast, and saturation of the images with a random factor in [-0.2, 0.2] and the hue with a random factor in [-0.1, 0.1]. The rotation operation randomly rotates each image of interest by an arbitrary angle.

\subsection{Preprocessing}
In addition to background removal, two popular preprocessing operations for fundus images are considered in this work, namely Graham processing \citep{graham} and contrast limited adaptive histogram equalization (CLAHE) \citep{clahe}. Both of them can alleviate the blur, low contrast, and inhomogeneous illumination issues that exist in the EyePACS dataset.

\begin{figure}[t]
    \centering
    \includegraphics[width=0.6\linewidth]{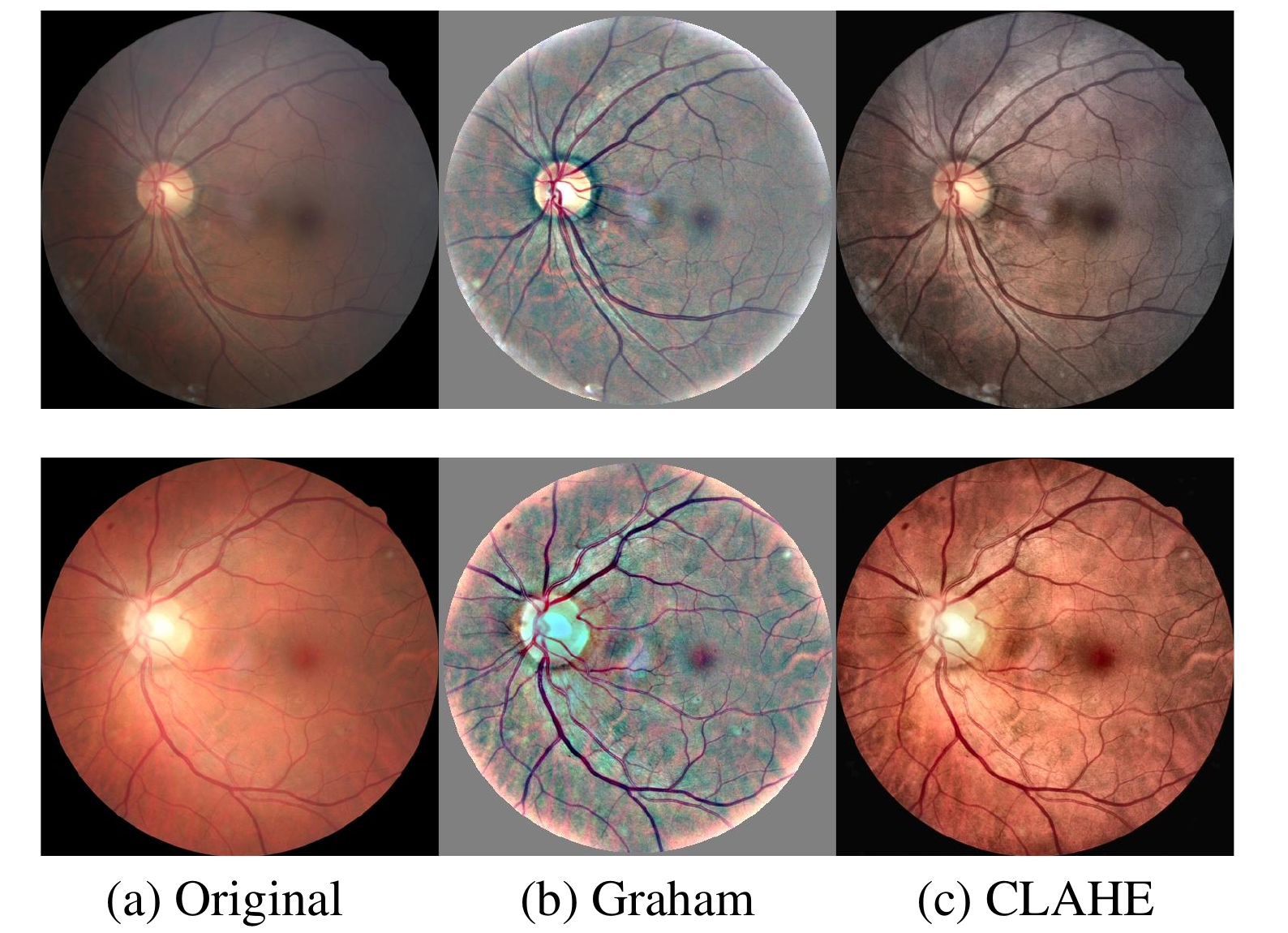}
    \caption{Representative enhanced fundus images using Graham processing and CLAHE. }
    \label{fig:pp}
\end{figure}

The Graham method was proposed by B. Graham the winner of the Kaggle DR grading competition. This preprocessing method has also been used in many previous works \citep{mining, twostage} to remove image variations due to different lighting conditions or imaging devices. Given a fundus image $\textbf{I}$, the processed image $\hat{\textbf{I}}$ after Graham is obtained by
\begin{equation}
    \hat{\textbf{I}} = \alpha \textbf{I} + \beta \textbf{G}(\theta) * \textbf{I} + \gamma,
\end{equation}
where $\textbf{G}(\theta)$ is a 2D Gaussian filter with a standard deviation $\theta$, $*$ is the convolution operator, and $\alpha, \beta, \gamma$ are weighting factors. Following \cite{twostage}, $\theta$, $\alpha$, $\beta$, and $\gamma$ are respectively set as 10, 4, -4, and 128. As shown in Fig.~\ref{fig:pp}, all images are normalized to be relatively consistent with each other and vessels as well as lesions are particularly highlighted after Graham processing.

CLAHE is a contrast enhancement method based on Histogram Equalization (HE) \citep{he}, which has also been widely used to process fundus images and has been suggested to be able to highlight lesions \citep{lesion, useclahe, useclahe2}. HE improves the image contrast by spreading out the most frequently-occurred intensity values in the histogram, but it amplifies noise as well. CLAHE was proposed to prevent an over-amplification of noise by clipping the histogram at a predefined value. Representative enhanced images via CLAHE are also illustrated in Fig.~\ref{fig:pp}.

\subsection{Sampling strategy}
As mentioned in section~\ref{sec:2.1}, EyePACS is an extremely imbalanced dataset. To address this problem, several sampling strategies \citep{tail,oO} for the training set have been proposed to rebalance the data distribution. Three commonly used sampling strategies are examined in this work: (1) instance-balanced sampling samples each data point with an equal probability. In this case, the class with more samples than the others can be dominant in the training phase, leading to model bias during testing; (2) class-balanced sampling first selects each class with an equal probability, and then uniformly samples data points from specific classes. In this way, samples in the minority classes are given more attention for better representation learning; (3) progressively-balanced sampling starts with class-balanced sampling and then exponentially moves to instance-balanced sampling. Please note that we follow the interpolation strategy adopted by \cite{oO} instead of the one presented by \cite{tail}, which linearly interpolates the sampling weight from instance-balanced sampling to class-balanced sampling. Specifically, the sampling weight in this work is defined as
\begin{equation}
    p^{\textrm{ \footnotesize PB}}_i(t) = \alpha^t p^{\textrm{\footnotesize CB}}_i + (1 - \alpha^t) p^{\textrm{\footnotesize IB}}_i,
\end{equation}
where $p^{\textrm{ \footnotesize PB}}, p^{\textrm{ \footnotesize CB}}$ and $p^{\textrm{ \footnotesize IB}}$ are sampling weights in progressively-balanced, class-balanced and instance-balanced sampling, $t$ indexes the training epoch and $\alpha$ is a hyper-parameter that controls the change rate.

\subsection{Prior knowledge}
\label{sec:pk}
\begin{figure}[t]
    \centering
    \includegraphics[width=\linewidth]{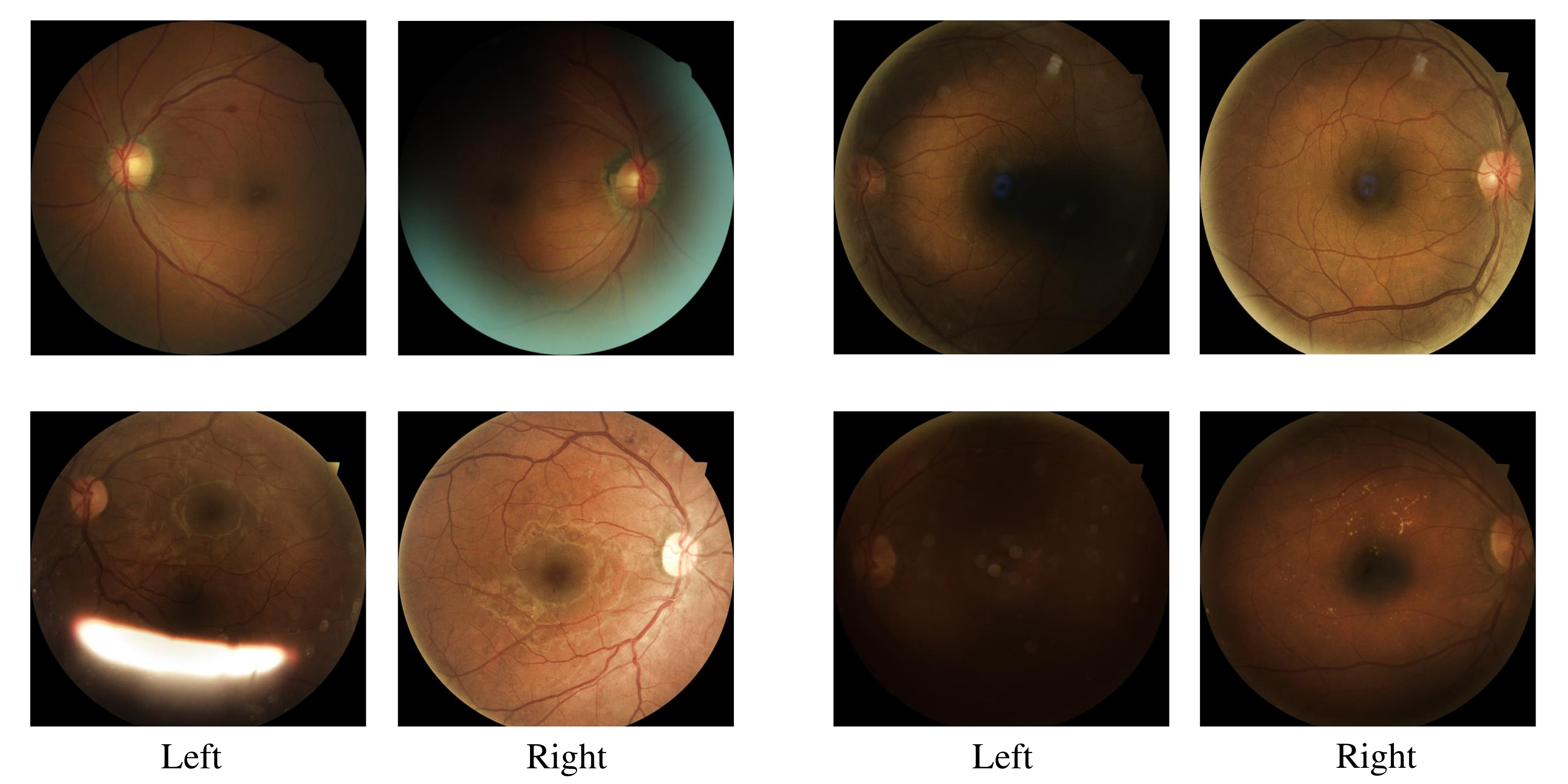}
    \caption{Representative eye pairs with different quality of the left and right fields.}
    \label{fig:pair}
\end{figure}

For medical image analysis, prior knowledge can significantly enhance the performance of deep learning frameworks. In the EyePACS dataset, both the left and right eyes of a patient are provided. Evidence shows that for more than 95\% the difference in the DR grade between the left and right eyes is no more than 1 \citep{zoom}.  Moreover, as demonstrated in Fig.~\ref{fig:pair}, the quality of the left and right fields of an eye pair may be different. And it is difficult to identify the grade of a fundus image with poor quality. In this case, information of the eye on the other side may greatly benefit the estimation of the grade of the poor one.

As such, to utilize the correlation between the two eyes, we concatenate the feature vectors of both eyes from the global average pooling layer of ResNet-50 and then input it into a paired feature fusion network. The network consists of 3 linear layers each followed by a 1D max-pooling layer with a stride of 2 and rectified linear unit (ReLU). Considering that the grading criterion for left and right eyes is the same, the feature fusion network only outputs the prediction for one eye and then changes the order of the two feature vectors during concatenation for the prediction of the other eye.

\subsection{Ensembling}
Ensemble methods \citep{ensemble} are widely used in data science competitions to achieve better performance. The variance in the predictions and the generalization errors can be considerably reduced by combining predictions from multiple models or inputs. However, ensembling too many models can be computationally expensive and the performance gains may diminish with the increasing number of models. To make our proposed pipeline generalizable, two simple ensemble methods are considered: 1) for the ensemble method that uses multiple models \citep{alexnet, multimodel}, we average the predictions from models trained with different random seeds. In this way, the datasets have different sampling orders and different data augmentation parameters to train each model, resulting in differently trained models for ensembling, 2) for the ensemble method that uses multiple views \citep{vgg, inception}, we first generate different image views via random flipping and rotation (test-time augmentation). Then these views including the original one are input into a single model to generate each view's DR grade score. We then use the averaged score as the finally predicted one.

\section{Experimental Results}
\label{sec:4}

\subsection{Influence of different input resolutions}
First, we study the influence of different input resolutions using the default setting specified in section~\ref{sec:2.1}. The experimental results are shown in Table~\ref{table1}. As suggested by the results, DR grading benefits from larger input resolutions at the cost of higher training and inference computational expenses. A significant performance improvement of 16.42\% in the test Kappa is obtained by increasing the resolution from $128 \times 128$ to $512 \times 512$. Increasing the resolution to $1024 \times 1024$ further improves the test Kappa by another 1.32\% but with a large computational cost increase of 64.84G floating-point operations (FLOPs). Considering the trade-off between performance and computational cost, the $512 \times 512$ input resolution is adopted for all our subsequent experiments.
\begin{table}[h]
    \centering
    \caption{DR grading performance with different input resolutions on EyePACS. Two GPUs are used to train the model with $1024 \times 1024$ input resolution due to the CUDA memory limitation.}
    \label{table1}
    \begin{tabular}{ccccc}  \hline
        Resolution & Training time & FLOPs & Validation Kappa & Test Kappa \\ 	\hline
        $128 \times 128$ & 1h 54m & 1.35G & 0.6535 & 0.6388 \\
        $256 \times 256$ & 2h 19m & 5.40G & 0.7563 & 0.7435 \\
        $512 \times 512$ & 5h 16m & 21.61G &  0.8054 & 0.8032 \\
        $768 \times 768$ & 11h 15m & 48.63G & 0.8176 & 0.8137 \\
        $1024 \times 1024$ & 11h 46m (2 GPUs) & 86.45G & 0.8187 & 0.8164 \\
        \hline
    \end{tabular}
\end{table}

\subsection{Influence of different objective functions}
We further evaluate the seven objective functions described in section~\ref{sec:loss}. We also evaluate the objective function by combining the Kappa loss and the cross-entropy loss \citep{Kappaloss}. All objective functions are observed to converge after 25 epochs of training. The validation and test Kappa scores for applying different loss functions are reported in Table~\ref{table2}. The results demonstrate the focal loss and the combination of the Kappa loss and the cross-entropy loss slightly improve the performance compared to the standard cross-entropy loss. The observation that using the Kappa loss alone makes the training process unstable and results in inferior performance is consistent with that reported in \cite{Kappaloss}. The MSE loss takes into account the distance between the prediction and the ground truth, yielding a 2.02\% improvement compared to the cross-entropy loss. It gives more penalties for outliers than the MAE loss and the smooth L1 loss, making itself have the highest validation and test Kappa among all the objective functions we consider.

\begin{figure}[t]
    \centering
    \includegraphics[width=0.7\linewidth]{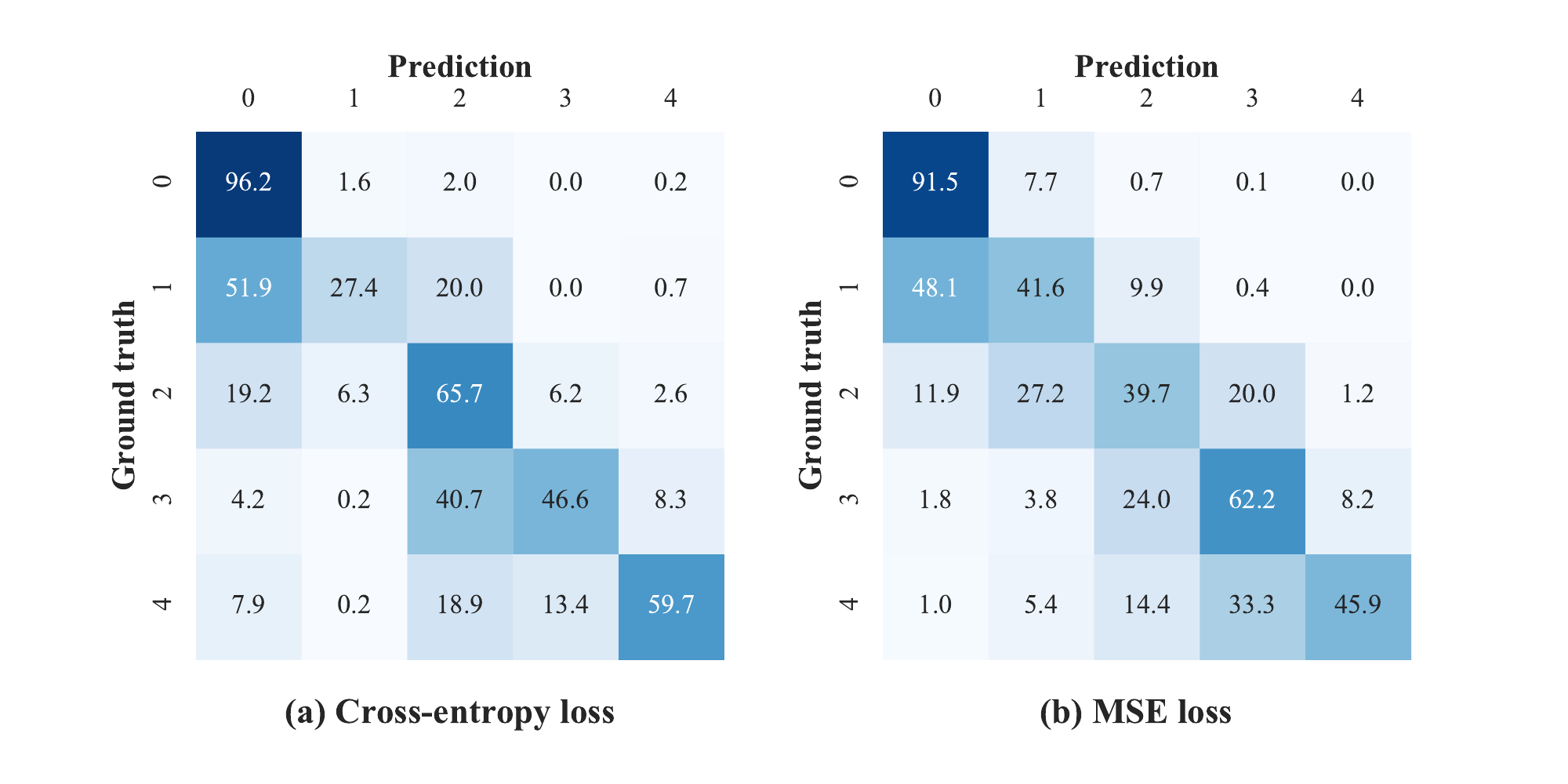}
    \caption{Confusion matrices from models respectively using the cross-entropy loss and the MSE loss as the objective function. All values in the confusion matrices are normalized.}
    \label{fig:dd}
\end{figure}

To demonstrate the influence of different objective functions on the distribution of predictions, we present the confusion matrics of the test set for the cross-entropy loss and the MSE loss in Fig.~\ref{fig:dd}. Considering the imbalanced distribution of different classes in EyePACS, we normalize the matrics by dividing each value by the sum of its corresponding row. As shown in Fig.~\ref{fig:dd}, although employing the MSE loss does not improve the performance of correctly discriminating each category, the prediction-versus-ground truth distance from using MSE is smaller than that from using cross-entropy (e.g. 7.9\% of proliferative DR images (Grade 4) are predicted to be normal when using the cross-entropy loss, while only 1.0\% when using the MSE loss). That is, the predictions from the model using the MSE loss as the objective function show more diagonal tendency compared to those using the cross-entropy loss, which contributes to the improvement in the Kappa metric. This diagonal tendency is important for DR grading in clinical practice because even if the diagnosis is wrong we expect our prediction to be at least close to the correct one.
\begin{table}[t]
    \centering
    \caption{DR grading performance of models using different objective functions on EyePACS . $\gamma$ is empirically set to be 2 for the focal loss.}
    \label{table2}
    \begin{tabular}{lcccc}  \hline
        Loss & Validation Kappa & Test Kappa \\ 	\hline
        Cross Entropy (CE) & 0.8054 & 0.8032 \\
        Focal ($\gamma$=2) & 0.8079 & 0.8059 \\
        Kappa & 0.7818 & 0.7775 \\
        Kappa + CE & 0.8047 & 0.8050 \\
        MAE & 0.7655 & 0.7679 \\
        Smooth L1 & 0.8094 & 0.8117 \\
        MSE & \textbf{0.8207} & \textbf{0.8235} \\
        \hline
    \end{tabular}
\end{table}

\subsection{Influence of different learning rate schedules}
Further on we study the influence of different learning rate schedules. All experiments are conducted using the baseline setting with the $512 \times 512$ input resolution and the MSE loss. The experimental results are shown in Table~\ref{table3}. The results demonstrate that except for the exponential decaying schedule, all schedules improve the Kappa on both the validation and test sets and the cosine decaying schedule gives the highest improvement of 0.32\% in the test Kappa. A plausible reason for the performance drop caused by the exponential decaying schedule is because the learning rate decreases too fast at the beginning of training. Therefore, the initial learning rate should be carefully tuned when the exponential decaying schedule is employed.

\begin{table}[t]
    \centering
    \caption{DR grading performance of models using different learning rate schedules on EyePACS. We set the initial learning rate to be 0.001 in all experiments. For the multiple-step decaying schedule, we decrease the learning rate by 0.1 at epoch 15 and epoch 20. For the exponential decaying schedule, we set the decay factor $\gamma$ to be 0.9.}
    \label{table3}
    \begin{tabular}{lcccc}  \hline
        Schedule & Validation Kappa & Test Kappa \\ 	\hline
        Constant & 0.8207 & 0.8235 \\
        Multiple Steps [15, 20] & 0.8297 & 0.8264 \\
        Exponential (p=0.9) & 0.8214 & 0.8185 \\
        Cosine & \textbf{0.8269} & \textbf{0.8267} \\
        \hline
    \end{tabular}
\end{table}

\subsection{Influence of different compositions of data augmentation}
We evaluate ResNet-50 with different compositions of data augmentation. In addition to flipping and rotation in the baseline setting, we consider random cropping, color jitter, and Krizhevsky color augmentation. We also evaluate the model trained without any data augmentation. All experiments are based on the best setting from previous evaluations. As shown in Table~\ref{table4}, even a simple composition of geometric data augmentation operations (the third row of Table~\ref{table4}) in the baseline setting can provide a significant improvement of 3.49\% on the test Kappa. Each data augmentation operation combined with flipping can improve the corresponding model's performance. However, the composition of all data augmentation operations considered in this work degrades the DR grading performance because too strong transformations may shift the distribution of the training data far away from the original one. Therefore, we do not simultaneously employ the two color transformations. The best test Kappa of 0.8310 is achieved by applying the composition of flipping, rotation, cropping, and color jitter for data augmentation during training. We adopt this composition in our following experiments.

\begin{table}[t]
    \centering
    \caption{DR grading performance of models using different compositions of data augmentation on EyePACS.}
    \label{table4}
    \begin{tabular}{ccccccc}  \hline
        Flipping &  Rotation & Cropping & Color jitter & Krizhevsky & Validation Kappa & Test Kappa \\ 	\hline
        & & & & & 0.7913 & 0.7923 \\
        \checkmark & & & & & 0.8124 & 0.8125 \\
        \checkmark & \checkmark & & & & 0.8258 & 0.8272 \\
        \checkmark & & \checkmark & & & 0.8194 & 0.8217 \\
        \checkmark & & & \checkmark & & 0.8129 & 0.8167 \\
        \checkmark & & & & \checkmark & 0.8082 & 0.8159 \\
        \checkmark & \checkmark & \checkmark & & & 0.8276 & 0.8247 \\
        \checkmark & \checkmark & \checkmark & \checkmark & & 0.8307 & \textbf{0.8310} \\
        \checkmark & \checkmark & \checkmark & & \checkmark & \textbf{0.8308} & 0.8277 \\
        \checkmark & \checkmark & \checkmark & \checkmark & \checkmark & 0.8247 & 0.8252 \\
        \hline
    \end{tabular}
\end{table}

\begin{figure}[t]
    \centering
    \includegraphics[width=\linewidth]{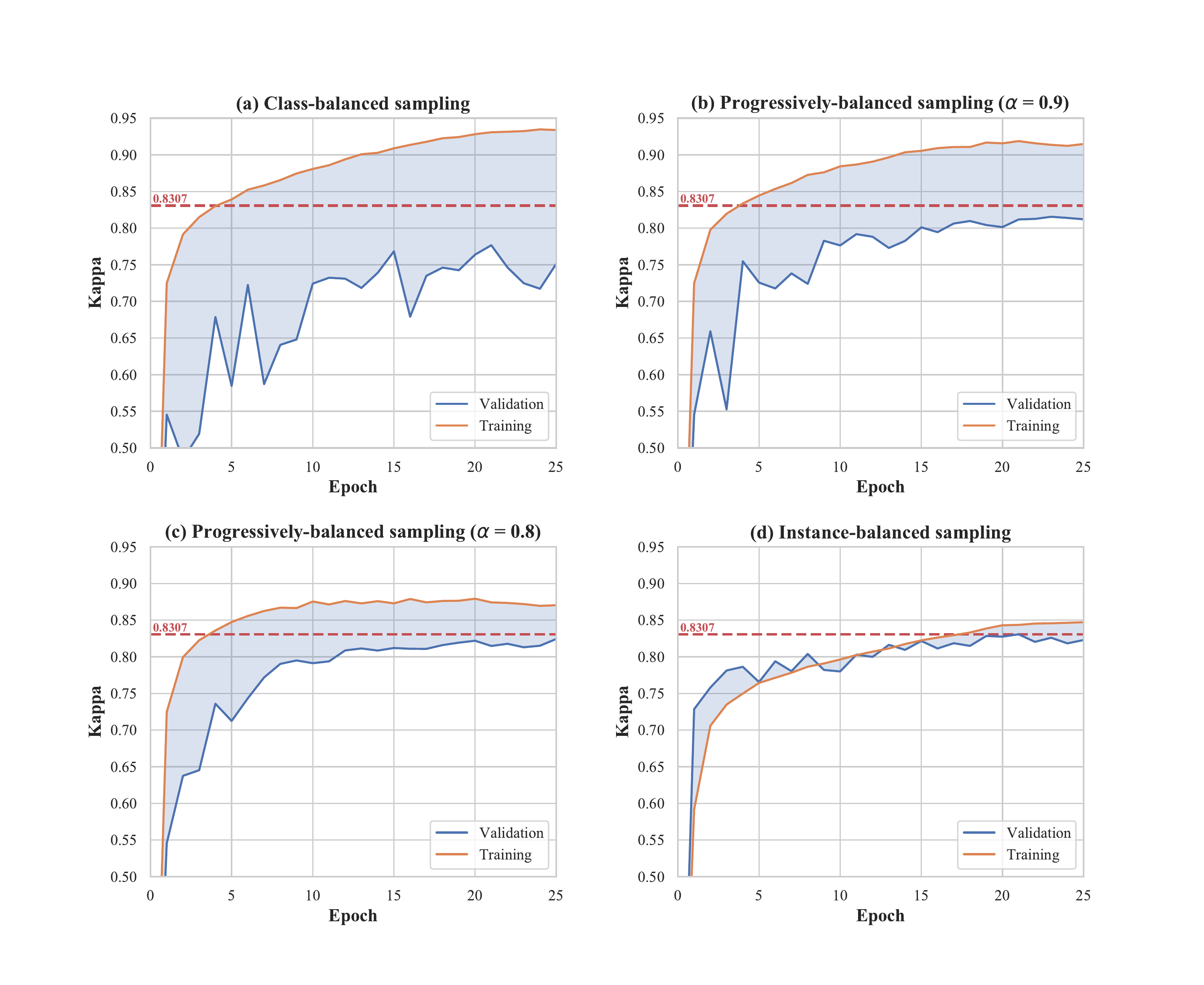}
    \caption{The performance of models using different sampling strategies for training. The dotted red line represents the best validation Kappa among these four experiments, which is achieved by instance-balanced sampling.}
    \label{fig:gap}
\end{figure}

\begin{table}[t]
    \centering
    \caption{DR grading performance on EyePACS with different preprocessing methods. Our default preprocessing setting consists of background removal and image resizing. The parameters used in the Graham method are set following \cite{twostage}. The clipping value and tile grid size of CLAHE are respectively set to be 3 and 8.}
    \label{preprocess}
    \begin{tabular}{lcccc}  \hline
        Preprocessing & Validation Kappa & Test Kappa \\ 	\hline
        Default & \textbf{0.8307} & \textbf{0.8310} \\
        Default + Graham \citep{graham} & 0.8262 & 0.8260 \\
        Default + CLAHE \citep{clahe} & 0.8243 & 0.8238 \\
        \hline
    \end{tabular}
\end{table}

\subsection{Influence of different preprocessing methods}
Two popular image enhancement methods are evaluated in our study, Graham processing and CLAHE. Both of them have been suggested to be beneficial for DR identification \citep{twostage, useclahe}. Although lesions become more recognizable with the application of the two preprocessing methods, they are not helpful for DR grading. As shown in Table~\ref{preprocess}, our framework with the Graham method achieves a 0.8227 test Kappa, which is lower than the default setting by about 0.5\%. Applying CLAHE also hurts the performance of our framework, decreasing the test Kappa by about 0.7\%. Unexpected noise and artifacts introduced by the preprocessing may be a cause of performance degradation in our experiments. As such, no image enhancement is applied in our following experiments.

\subsection{Influence of different sampling strategies}
Further, we concern about the influence of different sampling strategies. To alleviate the imbalance issue in EyePACS, class-balanced sampling and progressively-balanced sampling are conducted in the training phase. However, as illustrated in Fig.~\ref{fig:gap}, because we repeatedly sample data points from the minority classes at each epoch, overfitting results in poor performance on the validation set. The gap between the training Kappa and the validation Kappa increases as the probability of sampling the minority classes increases. Instance-balanced sampling, a strategy that we most commonly use, achieves the highest validation Kappa at the end of the training. A plausible reason for this result is that the class distribution of the training set is consistent with that of the validation set as well as those of real-world datasets. The class-based sampling strategies may be more effective in cases where the training set is imbalanced and the test set is balanced \citep{tail}.

\subsection{Influence of feature fusion of paired eyes}
We evaluate the improvement resulted from utilizing the correlation between the paired two eyes for DR grading. The best model from previous evaluations is fixed and adopted to generate feature vector of each fundus image. A simple paired feature fusion network described in section~\ref{sec:pk} is trained for 20 epochs with a batch size of 64. The learning rate is set to be 0.02 without any decaying schedule. As shown in Table~\ref{table6}, paired feature fusion improves the validation Kappa by 2.90\% and the test Kappa by 2.71\%, demonstrating the importance of the eye pair correlation to DR grading.

\subsection{Influence of different ensemble methods}

\begin{table}[t]
    \centering
    \caption{The performance of models with different ensemble methods on EyePACS.}
    \label{table5}
    \begin{tabular}{ccccc}
        \hline
        \multicolumn{1}{c}{\multirow{2}{*}{\# views / models}} & \multicolumn{2}{c}{Multiple views} & \multicolumn{2}{c}{Multiple models} \\ \cline{2-5}
        \multicolumn{1}{c}{}                                    & Validation Kappa         & Test Kappa         & Validation Kappa   & Test Kappa   \\
        \hline
        1                                                       & 0.8597                 & 0.8581                  & 0.8597           & 0.8581            \\
        2                                                       & 0.8611                 & 0.8593                  & 0.8622           & 0.8596            \\
        3                                                       & 0.8608                 & 0.8601                  & 0.8635           & 0.8615            \\
        5                                                       & 0.8607                 & 0.8609                  & 0.8644           & 0.8617            \\
        10                                                      & \textbf{0.8633}                 & 0.8603                  & \textbf{0.8660}           & \textbf{0.8631}            \\
        15                                                      & 0.8631                 & \textbf{0.8611}                  & 0.8653           & \textbf{0.8631}            \\
        \hline
    \end{tabular}
\end{table}

We also evaluate the impact of the number of input views for the ensemble method of multiple views and the number of models for the ensemble method of multiple models. The experimental results are tabulated in Table~\ref{table5}. We observe that as the number of models increases, both the test Kappa and the validation Kappa steadily increase. Unsurprisingly, the computational cost also monotonically increases with the number of ensembling. For the ensemble method that uses multiple models, the performance gain from increasing the number of models diminishes in the end and the best test Kappa is achieved by using 10 models.

\subsection{Comparison of the importance of all components}
\begin{figure}[t]
    \centering
    \includegraphics[width=0.6\linewidth]{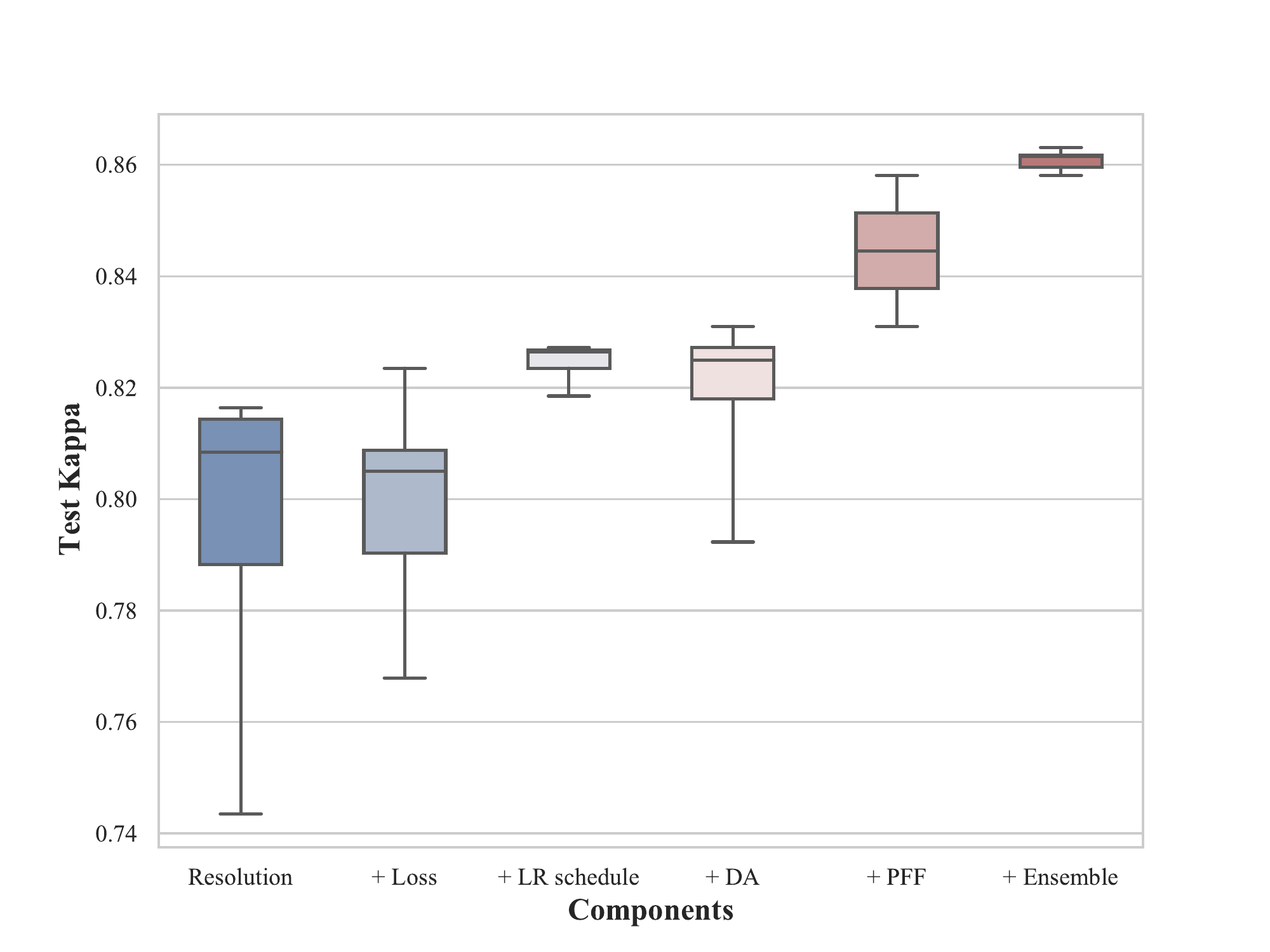}
    \caption{Box plots of the test Kappa of all experiments in this work. The experiments in each column are set up based on the best model considering all its left components. DA and PFF denote the experiment results of different compositions of data augmentation and applying paired feature fusion or not.}
    \label{fig:box}
\end{figure}

Finally, we investigate and compare the importance of all considered components in our DR grading task. We quantify the improvement from each component by applying them one by one, the results of which are shown in Table~\ref{table6}. We observe three significant improvements outstand from that table. First, increasing the input resolution from $224 \times 224$ to $512 \times 512$ gives the highest improvement of 5.97\%. Then, the choice of the MSE loss and utilization of the eye pair fusion respectively improve the test Kappa by another 2.03\% and 2.71\%. Additional improvements of 0.32\%, 0.43\%, and 0.5\% on the test Kappa are obtained by applying cosine decaying schedule, data augmentation, and ensemble (multiple models). Note that, the incremental results alone do not completely reflect the importance of different components. The baseline configuration may also affect the corresponding improvements. In Fig.~\ref{fig:box}, we present the ranges and standard deviations of all experiments in this work. If the range of a box is large, it indicates that the results of different choices of this component vary significantly. The top bar of the box represents the highest test Kappa that can be achieved by specifically refining the corresponding component. Obviously, a bad choice of either resolution, objective function or data augmentation may lead to a great performance drop. Applying a learning rate schedule and ensembling can both provide steady improvements but using different schedules or ensemble methods does not significantly change the DR grading result.

\begin{table}[t]
    \centering
    \caption{The performance of models on EyePACS for stacking refinements one by one. The first row is the result of the baseline we describe in section~\ref{sec:2.1}. HR, MSE, CD, DA, PFF, and ENS respectively denote the application of high resolution, MSE loss, cosine decaying schedule, data augmentation, paired feature fusion, and ensemble of multiple models.}
    \label{table6}
    \begin{tabular}{ccccccccc}  \hline
        HR &  MSE & CD & DA & PFF & ENS & Validation Kappa & Test Kappa & $\Delta$ test Kappa \\ 	\hline
        & & & & & & 0.7563 & 0.7435 & 0\% \\
        \checkmark & & & & & & 0.8054 & 0.8032 & +5.97\% \\
        \checkmark & \checkmark & & & & & 0.8207 & 0.8235 & +2.03\% \\
        \checkmark & \checkmark & \checkmark & & & & 0.8258 & 0.8267 & +0.32\% \\
        \checkmark & \checkmark & \checkmark & \checkmark & & & 0.8307 & 0.8310 & +0.43\% \\
        \checkmark & \checkmark	& \checkmark & \checkmark & \checkmark & & 0.8597 & 0.8581 & +2.71\% \\
        \checkmark & \checkmark & \checkmark & \checkmark & \checkmark & \checkmark& \textbf{0.8660} & \textbf{0.8631} & +0.50\% \\
        \hline
    \end{tabular}
\end{table}

\subsection{Comparison with state-of-the-art}
\begin{table}[t]
    \centering
    \caption{Comparisons with state-of-the-art methods on EyePACS with only image-level labels. Symbol `-' indicates the backbone of the method is designed by the corresponding authors. The results listed in the first three rows denote the top-3 entries on Kaggle's challenge.}
    \label{table7}
    \begin{tabular}{lccc}  \hline
        Method & Backbone & Test Kappa \\ 	\hline
        Min-Pooling & - & 0.8490 \\
        o\_O & - & 0.8450 \\
        RG & - & 0.8390 \\
        \hline
        Zoom-in Net \citep{zoom} & - & 0.8540 \\
        AFN \citep{afn} & - & 0.8590 \\
        CABNet \citep{cabnet} & ResNet-50 & 0.8456 \\
        Ours & ResNet-50 & 0.8581 \\
        Ours (ensemble) & ResNet-50 & \textbf{0.8631} \\
        \hline
    \end{tabular}
\end{table}

To assess the performance of our framework that incorporates the optimal set of all components investigated in this work, comparisons between the proposed method and previously-reported state-of-the-art ones without any utilization of additional datasets nor annotations are tabulated in Table~\ref{table7}. Our proposed method, without any fancy technique, outperforms previous state-of-the-art results by 0.91\% in terms of the test Kappa.

\begin{figure}[t]
    \centering
    \includegraphics[width=0.8\linewidth]{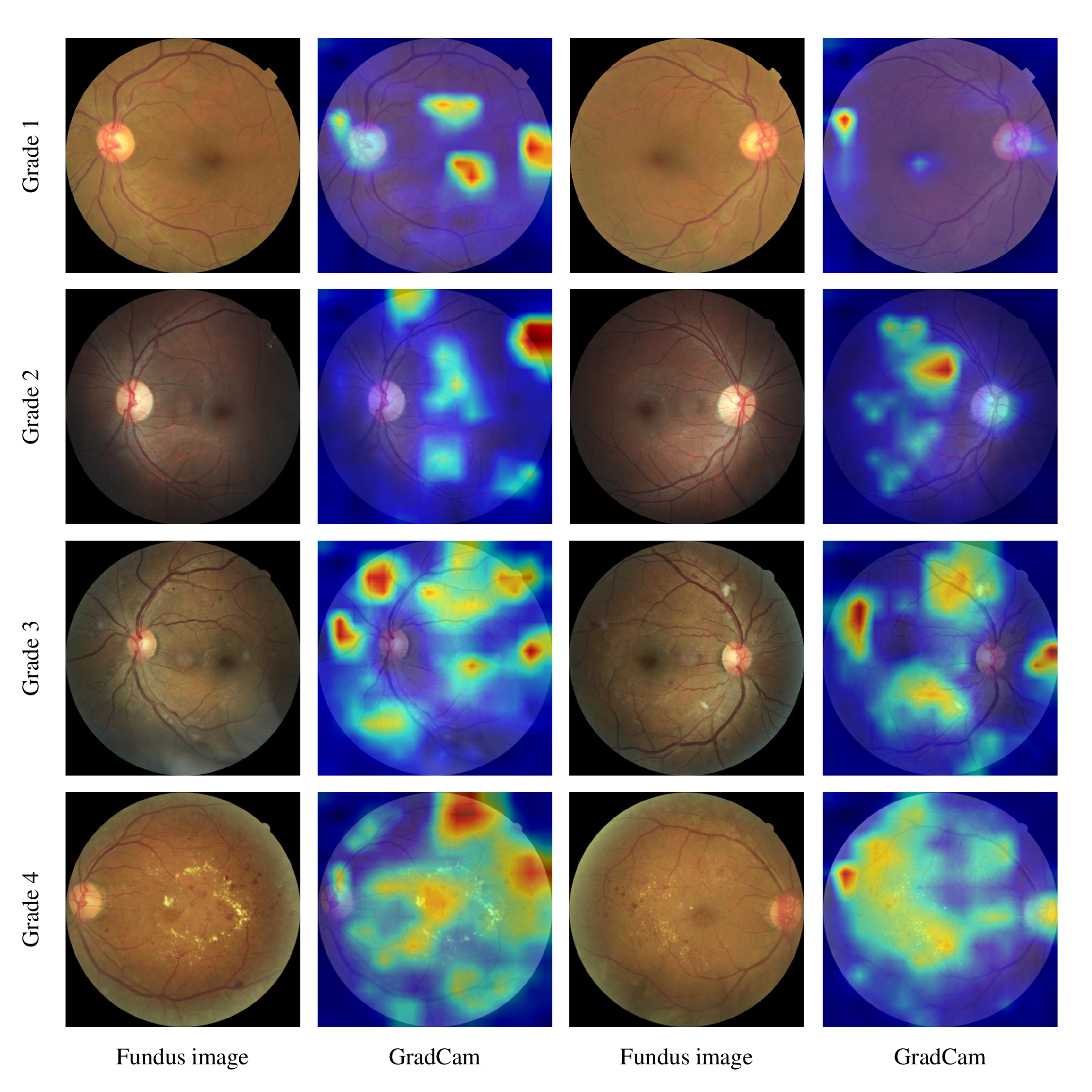}
    \caption{Visualization results from GradCAM. Representative eye pairs of four grades (mild DR, moderate DR, severe DR, and proliferate DR) are presented from top to bottom. The intensity of the heatmap indicates the importance of each pixel in the corresponding image for making the prediction.}
    \label{fig:cam}
\end{figure}

We then visualize our results using Grad-CAM \citep{gradcam}. As illustrated in Fig.~\ref{fig:cam}, representative results of four eye pairs corresponding to the four DR grades from 1 to 4 are provided. It reveals that our method's performance in DR grading may be a result of its ability to recognize different signs of DR, namely lesions. We observe that the region of the heatmap in a severe DR image is usually larger than that in a mild one because the amount of lesions to some degree reflects the DR grade and the lesions are what the network focuses on.

\subsection{Generalization ability of the refinements}

\begin{table}[t]
    \centering
    \caption{The DR grading performance on Messidor-2 and DDR datasets. Paired feature fusion is not feasible for the DDR dataset because eye pair information is not available for that dataset. HR, MSE, CD, DA, and PFF respectively denote the application of high resolution, MSE loss, cosine decaying schedule, data augmentation, and paired feature fusion.}
    \label{external}
    \begin{tabular}{ccccccccc}
        \hline
        \multirow{2}{*}{HR} & \multirow{2}{*}{MSE} & \multirow{2}{*}{CD} & \multirow{2}{*}{DA} & \multirow{2}{*}{PFF} & \multicolumn{2}{c}{Messidor-2}  & \multicolumn{2}{c}{DDR}         \\ \cline{6-9}
        &                      &                      &                     &                      & Test Kappa & $\Delta$ Kappa & Test Kappa & $\Delta$ Kappa \\ \hline
        &                      &                      &                     &                       & 0.7036        & 0\%             & 0.7680             & 0\%               \\
        $\checkmark$        &                      &                      &                     &                      & 0.7683        & +6.47\%         & 0.7870        & +1.90\%               \\
        $\checkmark$        & $\checkmark$         &                      &                     &                      & 0.7768        & +0.85\%         & 0.8000        & +1.30\%               \\
        $\checkmark$        & $\checkmark$         & $\checkmark$         &                     &                      & 0.7864        & +0.96\%         & 0.8056        & +0.56\%               \\
        $\checkmark$        & $\checkmark$         & $\checkmark$         & $\checkmark$        &                      & 0.7980        & +1.16\%         & \textbf{0.8326}        & +2.70\%               \\
        $\checkmark$        & $\checkmark$         & $\checkmark$         & $\checkmark$        & $\checkmark$         & \textbf{0.8205}             & +2.25\%               & -             & -               \\ \hline
    \end{tabular}
\end{table}

\begin{table}[t]
    \centering
    \caption{The DR grading performance on EyePACS using different network architectures. Underlining indicates that the improvement from the corresponding new component on that specific architecture is not consistent with that on ResNet-50. HR, MSE, CD, DA, and PFF respectively denote the application of high resolution, MSE loss, cosine decaying schedule, data augmentation, and paired feature fusion. MNet, D-121, RX-50, R-101, ViT-S, ViT-HS respectively denote MobileNet, DenseNet-121, ResNeXt-50, ResNet-101, small-scale Visual Transformer, small-scale hybrid Visual Transformer. $\kappa$ denotes Kappa score}
    \label{networks}
    \begin{tabular}{cccccccccccc}
        \hline
        &                       &                      &                      &                      & \multicolumn{6}{c}{Test Kappa}                                      &                                       \\ \cline{6-11}
        \multirow{-2}{*}{HR} & \multirow{-2}{*}{MSE} & \multirow{-2}{*}{CD} & \multirow{-2}{*}{DA} & \multirow{-2}{*}{PFF} & MNet & D-121 & RX-50 & R-101  & ViT-S & ViT-HS                   & \multirow{-2}{*}{Avg. $\Delta \kappa$} \\ \hline
        &                       &                      &                      &                      & 0.7517    & 0.7442      & 0.7395    & 0.7414 & 0.6797 & 0.7168                        & 0\%                                   \\
        $\checkmark$         &                       &                      &                      &                      & 0.7979    & 0.8046      & 0.8020    & 0.8075 & 0.7864 & 0.8073                       & +7.20\%                               \\
        $\checkmark$         & $\checkmark$          &                      &                      &                      & 0.8117    & 0.8158      & 0.8189    & 0.8228 & 0.8056 & 0.8256                       & +1.57\%                               \\
        $\checkmark$         & $\checkmark$          & $\checkmark$         &                      &                      & 0.8118    & 0.8255      & 0.8217    & \underline{0.8193} & \underline{0.8019} & 0.8257 & +0.09\%                               \\
        $\checkmark$         & $\checkmark$          & $\checkmark$         & $\checkmark$         &                      & 0.8226    & 0.8336      & 0.8362    & 0.8267 & 0.8215 & 0.8356                       & +1.17\%                               \\
        $\checkmark$         & $\checkmark$          & $\checkmark$         & $\checkmark$         & $\checkmark$         & \textbf{0.8515}         & \textbf{0.8558}           & \textbf{0.8566}         & \textbf{0.8528} & \textbf{0.836} & \textbf{0.8524}                            & +2.15\%                                     \\ \hline
    \end{tabular}
\end{table}

To evaluate the generalization ability of the proposed training setting refinements, two external retinal fundus datasets, Messidor-2 and DDR, are adopted to validate the models using the same training practices. As shown in Table \ref{external}, the improvements from each component on these two datasets are keeping in line with the results on EyePACS. Increasing the image resolution, applying the MSE loss and utilization of the eye pair fusion contribute significant improvements on the test Kappa scores. Incremental improvements are also observed from the learning rate schedule, data augmentation, and ensemble. Note that pair feature fusion is not utilized in the DDR dataset because eye pair labels are not available for that dataset. We observe that the key refinements we have identified for ResNet-50 based DR grading are shared across different datasets, such as the penalty to the distance between prediction and ground truth provided by the MSE loss is important for improving the Kappa metric. These consistent results demonstrate that the proposed training setting refinements can be generalized to other retinal datasets.

We also evaluate our proposed training settings on EyePACS using different backbones. Some popular model architectures are considered in this work, including a lightweight model MobileNet \citep{mobilenet}, a deeper model ResNet-101, and two ResNet variants DenseNet-121 \citep{dense}, ResNeXt-50 \citep{resnext}. We also look into recently-developed transformer based architectures, including small-scale Visual Transformer (ViT-S) \citep{dosovitskiy2020image} and small-scale hybrid Visual Transformer (ViT-HS) \citep{steiner2021train}. Because the architecture of visual transformers is largely different from that of CNNs, we adopt alternative training hyperparameters for our two ViT architectures following \cite{yu2021mil}. As shown in Table \ref{networks}, the consistent improvements from the investigated training practices, exerted to DR grading performance, reveal that the proposed practices can be generalized to different network architectures. We observe higher test Kappa scores for network architectures with more advanced designs or higher capacities. Notably, using cosine decaying as a learning rate schedule does not work well on ResNet-101 nor ViT-S. The reason may be due to the fact that our proposed refinements and configurations are determined empirically based on ResNet-50, and thus they may not necessarily be optimal for all other network architectures under consideration. Furthermore, we observe that cosine decaying is effective for all architectures without any other refinements, indicating that the order of stacking refinements may also affect the observed contribution of each component. With that being said, we show that our configurations can be a good starting point for tuning training strategies for DR grading.

\section{Discussion}
\label{sec:5}
Recently, deep learning methods have exhibited great performance on the DR grading task, but there is a trend that deep neural networks today become very large and highly sophisticated, making them difficult to be transferred and extended. Inspired by  \cite{miasurvey}, who states that the exact architecture is not the most important determinant in getting a good solution, we present a simple but effective framework without any dazzling design in the network architecture itself. Our proposed framework outperforms several state-of-the-art specifically-designed approaches tested on the EyePACS dataset. The promising performance of our proposed framework comes from the right choices of the input resolution, the objective function, the learning rate schedule, the composition of data augmentation, the utilization of the eye pair and the ensemble of multiple models. We also show that some popular techniques for fundus image-related tasks are not always beneficial for DR grading, such as image enhancement approaches and re-sampling strategies.

In this work, we focus on improving the DR grading performance of ResNet-50 on the EyePACS dataset. All refinements and configurations are determined empirically under that specific setting. Although we demonstrate that our refinements can generalize well to other network architectures and are robust across different datasets, our proposed solutions for DR grading may be still dependent on the property of the specific dataset of interest and the specific network of interest. In other words, our empirically-selected parameters may not be the best for other neural network architectures nor datasets. For example, the learning rate and its schedule need to be adjusted accordingly to identify the optimal solutions for frameworks using other types of neural networks as the backbones. The data augmentation composition may also need to be modified and the paired feature fusion strategy may be not always applicable for other DR grading datasets, such as the DDR dataset. Nevertheless, we show that our framework and the empirically-selected parameters can be a good starting point for the trial-and-error process during method design.

Our framework still has considerable room for improvement. In addition to the components we analyzed, there are other major components in deep learning based frameworks that are also worthy of being systematically investigated and refined. For example, regularization techniques, such as L1/L2 regularization and dropout \citep{srivastava2014dropout}, are essential to control the complexity of a model of interest to avoid overfitting, which may also affect the DR grading performance. In addition, how we combine different refinements and the order of stacking those different refinements may also have non-trivial impacts on the DR grading performance.

Recently, many specifically-designed components have been proposed to further improve the performance of deep learning-based methods using fundus images. Although they go beyond the scope of this work, those specifically-designed components may have great potential in enhancing the performance of DR grading. For example, the image quality is an important factor affecting the diagnoses of different ophthalmic diseases. Therefore, image quality enhancement \citep{cheng2021secret, zhao2019data} may serve as a preprocessing method to improve the DR grading performance. Another direction of improvement relates to the class imbalance issue of the EyePACS dataset. In this work, simple weighted resampling methods \citep{tail} are investigated, and the observed overfitting results indicate that these simple resampling methods are of limited help in improving the DR grading performance. Recently, a sophisticated sampling method, Balanced-MixUp \citep{galdran2021balanced}, has been proposed for imbalanced medical image classification tasks. In Balanced-MixUp, a more balanced training distribution is produced based on the MixUp regularization method \citep{zhang2017mixup}, and promising results have been reported on the DR grading task. Finally, more advanced data augmentation approaches, such as generative adversarial network based augmentation approaches \citep{zhou2020dr}, may be worthy of exploration to further boost the DR grading performance.

\section{Conclusion}
\label{sec:6}
In this work, we systematically investigate several important components in deep convolutional neural networks for improving the performance of ResNet-50 based DR grading. Specifically, the input resolution, objective function, learning rate schedule, data augmentation, preprocessing, data sampling strategy, prior knowledge, and ensemble method are looked into in our study. Extensive experiments on the publicly-available EyePACS dataset are conducted to evaluate the influence of different selections for each component. Finally, based on our findings, a simple yet effective framework for DR grading is proposed. The experimental results yielded from this study can be summarized as below.
\begin{itemize}
    \item We raise the ResNet-50 Kappa metric from 0.7435 to 0.8631 on the EyePACS dataset, outperforming other specially-designed DR grading methods. The generalization ability of the proposed training practices is successfully established on two external retinal fundus datasets and six other types of network architectures.
    \item Achieving state-of-the-art performance without any network architecture modification, we emphasize the importance of training setting refining in the development of deep learning based frameworks.
    \item Our codes and pre-trained model are publicly accessible at \href{https://github.com/YijinHuang/pytorch-classification}{https://github.com/YijinHuang/pytorch-classification}. We believe our simple yet effective framework can serve as a strong, standardized, and scalable baseline for further studies and developments of DR grading algorithms.
\end{itemize}

\section*{Acknowledgments}
The authors would like to thank Meng Li from Zhongshan Ophthalmic Centre of Sun Yat-sen University as well as Yue Zhang from the University of Hong Kong for their help on this work. This study was supported by the Shenzhen Basic Research Program (JCYJ20190809120205578); the National Natural Science Foundation of China (62071210); the Shenzhen Basic Research Program (JCYJ20200925153847004); the High-level University Fund (G02236002).

\bibliography{mybibfile}

\end{document}